\documentclass[aps,prb,reprint,superscriptaddress,floatfix]{revtex4-2}

\usepackage{siunitx}
\usepackage{physics}
\usepackage{chemformula}
\usepackage[inline]{enumitem}
\usepackage[hidelinks]{hyperref}

\usepackage{orcidlink}

\usepackage{enumitem,amssymb}
\newlist{todolist}{itemize}{2}
\setlist[todolist]{label=$\square$}
\usepackage{pifont}

\begin{document}


\title{Thorium-229 as a Phonomagnetometer}


\author{Martin Pimon\orcidlink{https://orcid.org/0000-0001-7784-463X}}
\email{martin.pimon@tuwien.ac.at}
\affiliation{Wolfgang Pauli Institute, Oskar-Morgenstern-Platz 1, 1090 Vienna, Austria}
\affiliation{Institute for Theoretical Physics, TU Wien, Wiedner Hauptstrasse 8-10/E136, 1040 Vienna, Austria}
\author{Andreas Grüneis\orcidlink{https://orcid.org/0000-0002-4984-7785}}
\affiliation{Institute for Theoretical Physics, TU Wien, Wiedner Hauptstrasse 8-10/E136, 1040 Vienna, Austria}
\author{Thorsten Schumm\orcidlink{https://orcid.org/0000-0002-1066-202X}}
\affiliation{Institute for Atomic and Subatomic Physics, TU Wien, Stadionallee 2, 1020 Vienna, Austria}
\author{Kjeld Beeks\orcidlink{https://orcid.org/0000-0002-8707-6723}}
\affiliation{Institute for Atomic and Subatomic Physics, TU Wien, Stadionallee 2, 1020 Vienna, Austria}


\date{\today}

\begin{abstract}
Thorium-229 possesses the only known low-energy nuclear transition suitable for spectroscopy with narrowband VUV lasers. While previous experiments have focused on application as a nuclear clock, this transition also offers a route to high-accuracy magnetometry. Experimental observations of magnetic fields generated by phonons carrying angular momentum remain inconclusive, highlighting the need for quantitative tests of the underlying physical mechanisms. In this article, we propose doping thorium-229 into a solid-state host to probe phonomagnetic fields in situ. We identify \ch{Na_2ThF_6}, a chiral stoichiometric thorium compound, as a promising host material and evaluate its spectroscopic sensitivity to magnetic interactions under two excitation schemes: driving degenerate phonon modes with a circularly polarized laser, and applying a temperature gradient. Density functional theory simulations combined with quantitative estimates suggest that the temperature-gradient scheme yields a magnetic signal that appears too weak to be measurable, while a circularly polarized, high-intensity THz/VUV pump--probe driven phonomagnetic response may approach the shot-noise-limited detection threshold. While experimental challenges remain, the unique suitability of thorium‑229 provides a testable pathway toward detecting phonomagnetic fields in a solid‑state platform.
\end{abstract}


\maketitle

\section{Introduction}
Lattice vibrations are conventionally treated as linear atomic displacements, yet under specific conditions, phonon modes can acquire angular momentum~\cite{Chiral_phonons_Jurasc_2025, Chiral_Phonons_Wang_2024}. In ionic crystals, displacements of atoms induce a polarization, described with the Born effective charge~\cite{gonze1997dynamical}, and the atomic motion in these modes forms a microscopic current loop, giving the phonon a magnetic moment~\cite{Orbital_magneti_Jurasc_2019, Anomalous_phono_Chaudh_2025}. The collective excitation of many such modes can thus generate a macroscopic magnetization and an associated magnetic field, the central signature of \emph{phonomagnetism}~\cite{Axial_phono_mag_Shabal_2025, Shabala2026Unifying}. The magnitude of this effect, however, remains contested. Reported phonon magnetic moments span several orders of magnitude~\cite{PhysRevResearch.4.013129, Chaudhary24Giant, Axial_phono_mag_Shabal_2025, Chen25Geometric, Chen25Gauge}, and the simple charge-current picture often underestimates measured values, pointing to additional electronic contributions~\cite{Effective_magne_Xiong_2022, Chaudhary24Giant}. Resolving this discrepancy is essential before phonomagnetism can be quantitatively understood; further measurements are urgently required.

The thorium-229 nucleus is unique among all known isotopes in possessing an exceptionally low-lying first excited state, the so-called isomer, situated only \SI{8.4}{\electronvolt} above the ground state~\cite{The229TVonDe2020, TheThorium229Beeks2021, TheThoriumIsoThirol2024, TheTickingOfTong2025}. The recent development of narrowband VUV laser sources via non-linear upconversion processes has enabled direct spectroscopic probing of this transition~\cite{tiedau2024laser,FrequencyRatioZhang2024, Laser_Mossbauer_Hiraki_2025, Frequency_repro_Ooi_T_2026, Continuous_wave_Morawe_2026, huang2026nuclear, lal2025continuous}. Its intrinsically narrow linewidth makes it an ideal next-generation frequency reference (``nuclear clock'')~\cite{NuclearLaserSPeik2003, PerformanceOfKazako2012, xiao2026continuous, de2026thorium}, and a sensitive probe of physics beyond the Standard Model~\cite{TheSearchForSafron2019, NuclearClocksPeik2021,beeks2025fine}. Two complementary experimental implementations are being pursued: trapped single ions, which offer the highest spectroscopic accuracy, and solid-state hosts, in which \ch{^{229}Th} is doped into a crystal in mesoscopic amounts. The solid-state approach yields a vastly superior signal-to-noise ratio owing to the large number of interrogated nuclei, at the cost of increased linewidth arising from interactions of the nuclear ensemble with the crystalline environment. These same interactions, however, encode detailed information on the local crystal fields, making the solid-state platform additionally attractive as a probe of the host's environment~\cite{Continuous_wave_Morawe_2026, Laser_Mossbauer_Hiraki_2025, Frequency_repro_Ooi_T_2026}

A stringent requirement on any host material is a band gap exceeding the $\sim \SI{8.4}{\electronvolt}$ transition energy, so that the crystal's electronic structure does not absorb the VUV photon; such wide gaps are found almost exclusively in fluorides~\cite{OpticalSpectroHehlen2013, A_spinless_crys_Morgan_2025}. A large band gap, however, implies a fully closed-shell electronic configuration with no unpaired electronic spins. Thus, the dominant magnetic perturbation reduces to dipolar coupling of the surrounding \ch{^{19}F} nuclei, orders of magnitude smaller than electrostatic sources of line splitting~\cite{PerformanceOfKazako2012, Frequency_repro_Ooi_T_2026}. This intrinsic magnetic quietness renders the system attractive for magnetometry, since an internal field can produce resolvable line shifts against a narrow baseline. Despite this prospect, the \ch{^{229}Th} solid-state transition has not yet been employed as a magnetometer.

In this work, we propose using this transition to measure the internal magnetic fields generated by phonomagnetism. Beyond its diamagnetic properties, the solid-state \ch{^{229}Th} clock is particularly well suited to this task for several reasons. The nucleus is embedded in the lattice rather than introduced as an external probe, so it reports on the same phonon environment that generates the field of interest with point-like magnetic-dipole coupling. Because the transition is optically driven and read out directly (a task made tractable by knowledge of the local crystal-field environment), no external field is required, avoiding the risk of masking a weak phonomagnetic signal with a much larger applied field. Finally, the transition's narrow linewidth, further stabilized by the long isomer lifetime, translates directly into a high sensitivity to small frequency shifts.

By contrast, established techniques are not well suited to measuring phonomagnetism. SQUIDs and NV-center magnetometers are external probes, limited by $1/r^3$ dipolar falloff and poor bulk resolution. Magneto-optical Kerr-effect microscopy, recently used to detect circularly polarized phonons~\cite{han2026coupled}, operates in reflection and is restricted to the optical penetration depth. NMR probes embedded nuclei but requires a large polarizing field that would overwhelm the much weaker phonomagnetic signal. Magnetic fields at nuclei can also be measured through line splitting in Mössbauer spectra~\cite{evans2017magnetic} or quantum-beating patterns in nuclear resonant scattering (NRS)~\cite{rohlsberger2004nuclear} on other established nuclear isomers. However, using these approaches to measure phonomagnetism, which is expected to cause a $< \si{\kilo\hertz}$ splitting on a $< \si{\nano\second}$ timescale, is challenging. NRS cannot resolve $< \si{\kilo\hertz}$ splittings, since the corresponding beat period would extend into the millisecond range. Mössbauer spectroscopy of $^{67}$Zn has been used to resolve $\sim \SI{100}{\hertz}$ shifts on a $\sim \SI{20}{\kilo \hertz}$ linewidth~\cite{potzel199267zn}, but this technique cannot exploit pump--probe methods that take advantage of the short phonon lifetime. Finally, we note that the Born effective charges of the fluoride hosts under consideration are not exceptionally large compared to materials with soft polar phonon modes, such as \ch{SrTiO3}~\cite{choi2024real}. The case for \ch{^{229}Th} as a phonomagnetic probe therefore rests not on an enhanced phonon-induced magnetic field, but on the combination of a low magnetic background, embedded and non-perturbative detection, and sensitivity of the clock transition.


\section{Theory}

\subsection{Phonomagnetic Field}

We begin with the definition of the Born effective charge tensor, $eZ^*_{a,ij} = \pdv{P_i}{u_{a,j}}$, where $i$ is the direction of the polarization with respect to a rigid displacement of atom $a$ in the direction of $j$. This allows us to express the ionic contribution to the displacement current density as
$\vb{j}_a(\vb{r}) = \pdv{\vb{P}_a}{t} = e \vb{Z}_a^* \dot{\vb{u}}_a \delta^3 [\vb{r}-(\vb{r}_a+\vb{u}_a)],$
where $\vb{u}_a$ is the displacement of atom $a$ due to lattice vibrations. Sufficiently far from the microscopic current loops generated by this motion, the associated orbital magnetic moment is $\vb{M}_a = \frac{1}{2} \int_V (\vb{u}_a \times \vb{j}_a) \,\dd^3 \vb{r}.$ Because $\vb{j}_a$ is spatially localized around atom $a$, this integral reduces to
\begin{equation}
    \vb{M}_a = \frac{e}{2} [\vb{u}_a \times (\vb{Z}_a^* \dot{\vb{u}}_a)].
    \label{eq:magmom1}
\end{equation}

To evaluate this expression, we adopt a classical picture in which the displacement of atom $a$ oscillates as $\vb{u}_a(t) = \frac{1}{\sqrt{m_a}} \Re[ A \vb{e}_a e^{-i \omega t}]$ where $\vb{e}_a$ is the mass-normalized polarization vector of the vibrational mode, satisfying $\sum_a \abs{\vb{e}_a}^2 = 1$. Decomposing $\vb{e}_a$ into real and imaginary parts, $\vb{e}_a = \vb{e}'_a+i \vb{e}''_a$, we obtain $\vb{u}_a(t)  = A(\vb{e}'_a \cos \omega t + \vb{e}''_a \sin \omega t)$, with the corresponding velocity $\dot{\vb{u}}_a(t)  = A \omega (- \vb{e}'_a \sin \omega t + \vb{e}''_a \cos \omega t)$. Substituting these expressions into \autoref{eq:magmom1} and time-averaging, using $\langle \cos^2 \omega t \rangle  = \langle \sin^2 \omega t \rangle=\frac{1}{2}$ and $\langle \sin \omega t \cos \omega t \rangle = 0$, yields, after some algebra,
\begin{equation}
    \vb{M}_a = \frac{e \hbar}{2 m_a} \Im[\vb{e}_a^* \times (\vb{Z}^*_a \cdot \vb{e}_a)],
    \label{eq:magmom2}
\end{equation}
where we have made the substitution $A^2 \rightarrow 2\hbar/\omega$ to express the result in terms of the vibrational quantum. This expression holds for any atomic trajectory and fully accounts for the anisotropy of the $3\times3$ Born effective charge tensor $\vb{Z}^*_a$. In the special case where $\vb{Z}^*_a$ is isotropic and can be treated as a scalar $Z^*_a$, \autoref{eq:magmom2} reduces to the familiar form
$\vb{M}_a = \frac{e Z^*_a}{2 m_a} \vb{L}_a,$
where $\vb{L}_a = m_a (\vb{u}_a \times \dot{\vb{u}}_a) = \hbar \Im[\vb{e}_a^* \times \vb{e}_a]$ is the angular momentum of atom $a$.


Finally, we consider how these localized orbital magnetic moments influence a nearby probe site. A set of magnetic dipoles $\vb{M}_a$, located at positions $\vb{r}_a$ relative to the thorium site (taken as the origin), produces a magnetic field there given by
\begin{equation}
\vb{B}(0) = \frac{\mu_0}{4\pi} \sum_a \left( \frac{3\vb{r}_a(\vb{M}_a \cdot \vb{r}_a)}{r_a^5} - \frac{\vb{M}_a}{r_a^3} \right).
\label{eq:Bfield}
\end{equation}
In our proposal, these moments arise from phonon-induced microscopic current loops on the surrounding atoms in large-gap ionic crystals ($\mu_r \approx 1)$, and we use this expression to compute the resulting magnetic field at the thorium site.

\subsection{Nuclear Energy Shifts}

The nuclear energy levels of an isotope embedded in a solid-state host are perturbed by the local electric and magnetic fields at the nuclear site. Here, we outline the four leading-order interactions that determine the resulting spectrum~\cite{Mossbauer_Spect_Greenw_1971, ConstrainingThReller2010, PerformanceOfKazako2012, Sup229SupTDessov2014}:
\begin{enumerate}
    \item The isomer shift $H_{E0} = e Z \rho_{\text{el}}(0) \langle \Delta R^2 \rangle / 6 \varepsilon_0$, arises from the contact interaction between the electronic charge density at the nucleus $\rho_{\text{el}}$, and the change in mean-square nuclear charge radius across the nuclear transition, $\langle \Delta R^2 \rangle = \SI{0.0107}{\square \femto \metre}$~\cite{Continuous_wave_Morawe_2026}.
    \item Quadrupolar splitting, $H_{E2} = Q V_{zz}/4I(2I-1) \times \left(3I_z^2 - \vb{I}^2 + \eta \left[I_x^2 - I_y^2 \right] \right)$,  arises from the coupling between the nuclear quadrupole moment $Q$, and the electric field gradient (EFG) tensor $V_{ij} = \left. \pdv[2]{V}{x_i}{x_j} \right|_0$. The quadrupole moments of the ground and excited nuclear states are $Q_{\text{g}} = \SI{3.11}{e\barn}$ and $Q_{\text{is}} = \SI{1.77}{e\barn}$, respectively~\cite{FrequencyRatioZhang2024}.
    \item Magnetic splitting, $H_{M1} = - (\mu / \hbar I) \vb{I} \cdot \vb{B}$, results from the Zeeman interaction between the nuclear magnetic moment $\mu$ and the local magnetic field $B$. For thorium-229, $\mu_{\text{g}} = \SI{0.36}{\mu_N}$ and $\mu_{\text{is}} = \SI{-0.37}{\mu_N}$~\cite{TheThorium229Beeks2021}.
    \item The second-order Doppler shift $H_D = -E_{\gamma}  \langle v^2 \rangle / 2c^2$ arises from the time-averaged kinetic energy of the thermally oscillating nucleus.
\end{enumerate}
Expressed in the principal-axis system of the electric field gradient, the total Hamiltonian is the sum of these contributions:
\begin{equation}
    H = H_{E0} + H_{E2} + H_{M1} + H_D.
\end{equation}

While all four terms are formally present, their magnitudes differ substantially for the \ch{^{229}Th} nucleus in solid-state environments. The second-order Doppler shift is negligible, as the heavy thorium nucleus has a small mean-square velocity even at elevated temperatures, and $E_\gamma/c^2$ further suppresses $H_D$ well below the other contributions. The quadrupolar splitting $H_{E2}$ dominates the hyperfine spectrum, reaching tens to hundreds of MHz even for modest electric field gradients, while the isomer shift $H_{E0}$ is comparatively small, amounting to a few MHz between distinct local environments in \ch{Th:CaF_2}~\cite{Continuous_wave_Morawe_2026}. For magnetic interactions $H_{M1}$, in the absence of unpaired electrons or an external field, only the nuclear magnetic moment of the surrounding \textsuperscript{19}F contributes ($\sim \SIrange{0.1}{0.5}{\milli \tesla}$). The broadening due to this nuclear spin bath has not been isolated experimentally to date, but it is estimated to lie between \SIrange{0.5}{1.5}{\kilo\hertz}~\cite{PerformanceOfKazako2012}.

\subsection{Computational Methods}

All numerical calculations were performed within the framework of density functional theory (DFT), as implemented in the Vienna Ab initio Simulation Package (VASP)~\cite{vasp1, vasp2, vasp3, vasp4}, using the PBEsol exchange-correlation functional~\cite{pbesol}. Second- and third-order interatomic force constants were computed from VASP-derived forces using the phonopy~\cite{phonopy1, phonopy2} and phono3py~\cite{phonopy1, phono3py} packages within the displacement-supercell approach. Material-specific computational parameters, including supercell sizes, $k$-point sampling, and energy cutoffs, are given alongside the corresponding results in the Proposed Measurements section.

\section{Proposed Measurements}
\label{sec:measurements}
\subsection{Material Selection}
We now turn to experimental implementation and consider two strategies for generating phonomagnetic fields:
\begin{enumerate}
    \item driving degenerate modes at the $\Gamma$-point with a THz laser~\cite{Orbital_magneti_Jurasc_2019}, and
    \item applying a temperature gradient in a non-centrosymmetric material belonging to a gyrotropic point group~\cite{PhysRevResearch.2.012073, PhysRevB.108.134307}, which admits axial rank-two pseudotensors and thus a phonon-driven magnetic flux~\cite{Hamada2018}.
\end{enumerate}
We focus on thorium-229 as a dopant in stoichiometric crystals to minimize strain-induced inhomogeneous broadening~\cite{Frequency_repro_Ooi_T_2026}. To identify suitable host candidates, we queried the Materials Project structure database~\cite{Jain2013, Ong2012b, Ong_2015} for large-gap, stoichiometric thorium compounds. Although this database is extensive, its DFT-calculated band gaps are systematically underestimated~\cite{DensityFunctioPerdew2009}, so we imposed a lower bound on the queried band gap well below the isomeric energy. Since \textcite{RadiativeDecayPineda2025} were unable to observe the thorium transition in a \ch{SiO_2} host, we speculate that $\ch{O} \rightarrow \ch{Th}$ charge-transfer excitations produce band gaps smaller than the isomeric transition energy. Among the stoichiometric thorium oxide compounds in the database, the largest calculated band gap is \SI{5.41}{\electronvolt}, which we adopt as our lower cutoff.

Only 17 thorium-stoichiometric crystals exceed this threshold. Of these, five are neither gyrotropic nor host degenerate modes, six are gyrotropic, and nine feature degenerate modes. To enable a meaningful comparison between the two experimental approaches, we further required the crystal to be both gyrotropic and to host degenerate modes, narrowing the candidates to three. In each of these three crystals, the degenerate modes transform as $(x, y)$, defining the circular polarization plane of the THz laser and yielding a magnetic moment aligned with $z$. However, only one material, \ch{Na_2ThF6}, with point group $D_3$, permits a nonzero phonomagnetic response to a temperature gradient along $z$~\cite{Aroyo:xo5013, PhysRevResearch.2.012073}, thereby enabling a direct comparison between both approaches.

\begin{figure}[htbp!]
    \centering
    \includegraphics[width=1.0\linewidth]{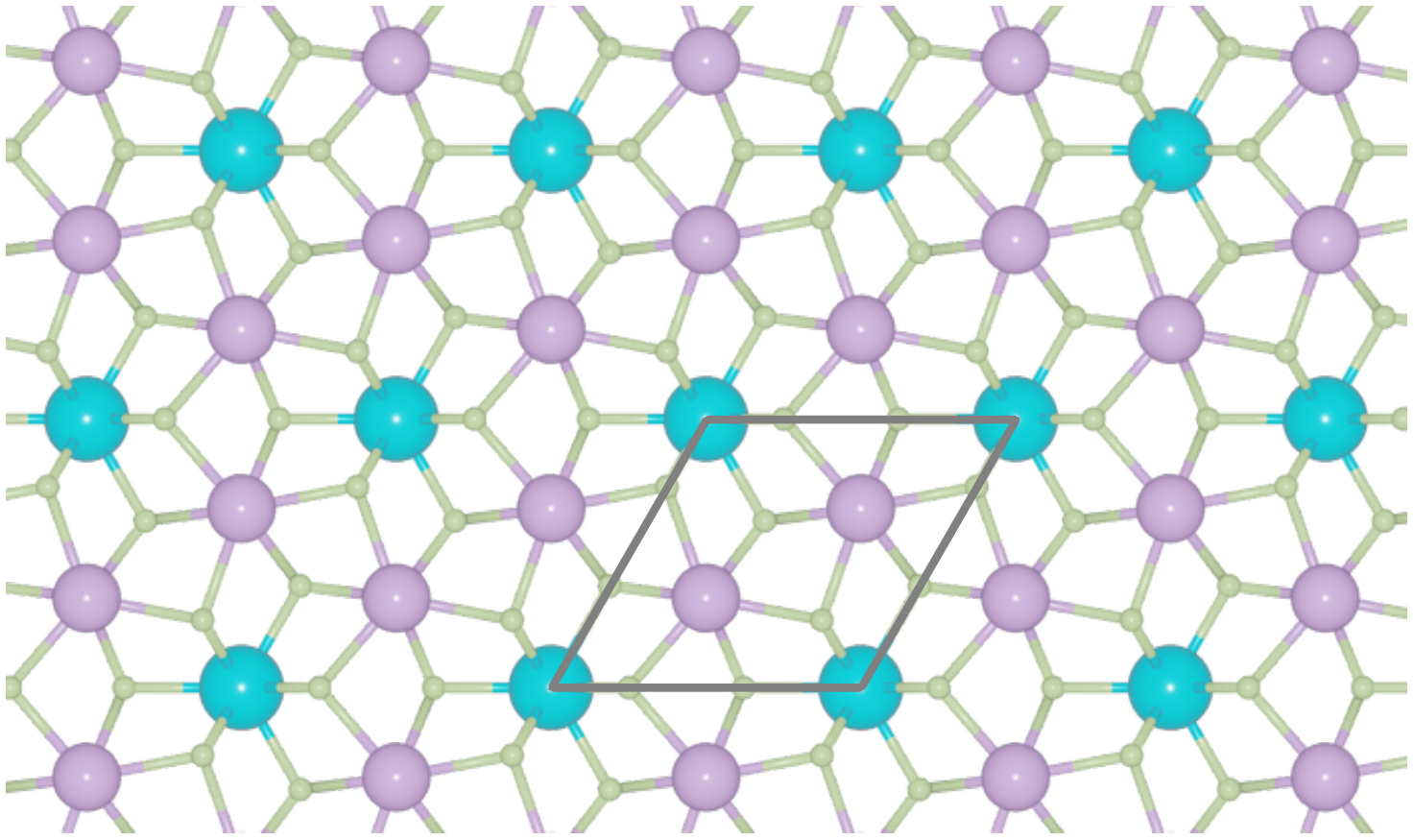}
    \caption{Crystal structure of \ch{Na_2ThF_6} viewed from the crystal $z$-axis. Th atoms are blue, Na atoms are purple and F atoms are lime. The primitive cell is indicated with a gray line.}
    \label{fig:structure}
\end{figure}

\ch{Na_2ThF_6} crystallizes in the chiral, non-enantiomorphic, trigonal $P321$ Sohncke space group (no.~150)~\cite{Chirality_in_th_Fecher, Chirality_deter_Valent_2022, Bousquet_2025} (see \autoref{fig:structure}), confirmed stable between \SIrange{100}{954}{\kelvin} and \SIrange{0.0001}{6.4}{\giga\pascal}~\cite{Crystal_structu_Grzech_2007}, with a melting point of \SI{981}{\kelvin} at standard conditions~\cite{Investigation_o_Emelya_1956}. The primitive unit cell contains one formula unit, with Th at Wyckoff position 1a, Na at 2d, and F at the 3e and 3f sites~\cite{Experimental_an_Schreu_2021}, yielding 27 total degrees of freedom. The three acoustic modes transform as $A_2 + E$, while the remaining 24 optical modes decompose as $3A_1 + 5A_2 + 8E$; hence, the crystal possesses eight symmetry-required, doubly-degenerate $E$ phonon branches at the zone center~\cite{TEIXEIRA2003159}.

\textcite{TEIXEIRA2003159} grew \ch{Na_2ThF_6} single crystals of several \si{\cubic \centi \meter} and investigated the vibrational spectrum experimentally; a theoretical study was done by \textcite{First_principle_Togo}. Several authors have proposed \ch{Na_2ThF_6} as a candidate host for a solid-state nuclear clock~\cite{rellergert2010progress, OpticalSpectroHehlen2013, Creation_of_inv_Tkalya_2013, InvestigationOEllis2014, SolidStateMmlXuHa2023}. \textcite{rellergert2010progress} recorded its VUV fluorescence spectrum and concluded it was transparent and suitable for clock operation, but they expressed concern over radiation damage under intense VUV light. \textcite{InvestigationOEllis2014} and \textcite{SolidStateMmlXuHa2023} determined a fundamental band gap of around \SI{11}{\electronvolt} theoretically, using the $G_0W_0$ approximation~\cite{New_Method_for_Hedin_1965}.

\begin{figure}[htpb!]
    \centering
     \includegraphics[width=1.0\linewidth]{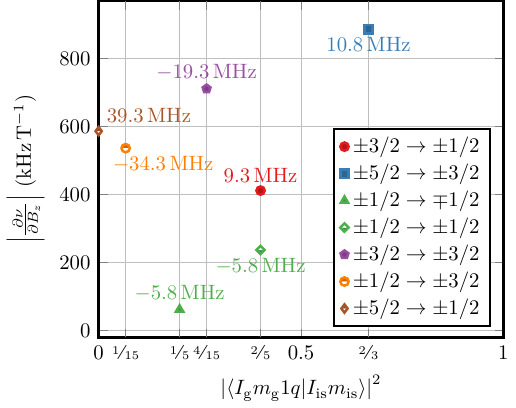}
    \caption{Nuclear transition line sensitivity to magnetic fields. The transition probability corresponds to the Clebsch-Gordan coefficient coupling the photon angular momentum to the nuclear ground- and excited-state sublevels. The numerical values indicate the frequency shifts of the sublevel transitions relative to the unsplit ground- to excited-state transition, calculated for \ch{Na_2ThF_6}.}
    \label{fig:sensitivityB}
\end{figure}

In this study, we optimized the lattice vectors and ionic positions using an energy cutoff of \SI{675}{\electronvolt} on a $5 \times 5 \times 7$ k-point mesh, with the \texttt{F}, \texttt{Na\_pv}, and \texttt{Th} PAW pseudopotentials, until the largest force component fell below \SI{1}{\milli\electronvolt\per\angstrom} and the largest stress-tensor component fell below \SI{0.1}{\kilo\bar}. The resulting lattice parameters were $a = b = \SI{5.97}{\angstrom}$ and $c = \SI{3.81}{\angstrom}$. We calculated an electric field gradient of $V_{zz} = \SI{-13.41}{\volt\per\square\angstrom}$ with a symmetry-enforced asymmetry parameter $\eta = 0$, the principal $z$-axis coinciding with the crystal $z$-axis. We further examined the sensitivity of the thorium-229 sublevel transitions to magnetic perturbations, modeled as an effective flux density aligned with the EFG's principal $z$-axis; results are shown in \autoref{fig:sensitivityB}.
Conveniently, the $m_{\text{g}} \rightarrow m_{\text{is}} = \pm 5/2 \rightarrow \pm 3/2$ transition with the strongest magnetic-field sensitivity also has the highest transition probability, as given by the Clebsch-Gordan coefficient $\abs{\braket{I_{\text{g}} m_{\text{g}} 1 q}{I_{\text{is}} m_{\text{is}}}}^2 = 0.6$, where $I_{\text{g}} = 5/2$ and $I_{\text{is}} = 3/2$ are the nuclear ground- and excited-state spins; the vanishing asymmetry parameter ($\eta = 0$) precludes state mixing.


\subsection{Circularly Polarized Laser Drive}


The symmetry of the Th and Na sites restricts their motion to the $x$-$y$ plane, whereas the lower symmetry at the F sites permits a small but nonzero $z$-component in their oscillation. We denote the polarization vectors of both degenerate modes with $\vb{e}^{(1)}_a$ and $\vb{e}^{(2)}_a$, and construct a circular basis $\vb{e}^{\pm}_a = (\vb{e}^{(1)}_a \pm i \vb{e}^{(2)}_a) / \sqrt{2}$. In this basis, the atomic displacement and velocity read $\vb{u}_a = \sqrt{\frac{\hbar}{2 m_a \omega}}\sum_{\pm} \vb{e}^{\pm}_a (a_{\pm} e^{i \omega t} + a^{\dagger}_{\pm} e^{-i \omega t})$ and $\dot{\vb{u}}_a = i \sqrt{\frac{\hbar \omega }{2 m_a}}\sum_{\pm} \vb{e}^{\pm}_a (a_{\pm} e^{i \omega t} - a^{\dagger}_{\pm} e^{-i \omega t})$ with $a_{\pm} = (a_1 \mp i a_2)/\sqrt{2}$. To compute the magnetic moment, we start from \autoref{eq:magmom1}, average over time, discard the rapidly oscillating terms, and retain only the number operators $\langle a_{\pm}^{\dagger} a_{\pm} \rangle = n_{\pm}$. This yields our final result:
\begin{equation}
    \langle \vb{M}_a \rangle = \frac{e \hbar}{2 m_a} \Im[(\vb{e}^{\pm}_a)^{*} \times (\vb{Z}^*_a \cdot \vb{e}^{\pm}_a)] (n_+ - n_-).
    \label{eq:laser_magmom}
\end{equation}

Estimating the magnitude of the phonon magnetic moment, and hence the induced magnetic field, requires knowledge of the expected occupation-number difference between the two polarization helicities. To obtain an upper bound, we model the system as a classical, driven, damped harmonic oscillator $\ddot{Q}_{\pm} + 2 \gamma_{\nu}(T) \dot{Q}_{\pm} + \omega_{\nu}^2 Q_{\pm} = e E_{\pm} e^{i\omega t} Z_{\text{eff}}^*$, where $Z_{\text{eff}} = \sum_a \frac{1}{\sqrt{m_a}} (\vb{e}^{\pm}_a)^{\dagger} \cdot \vb{Z}^*_a \cdot \hat{\vb{\epsilon}}^{\pm}$ and $\gamma_{\nu}(T)$ is the damping rate, corresponding to the phonon linewidth (HWHM). We make the following simplifying assumptions: the THz laser is perfectly circularly polarized, $\hat{\vb{\epsilon}}^{\pm} = (\hat{\vb{x}} \pm i \hat{\vb{y}})/\sqrt{2}$; $E_+$ drives only $Q_+$, and vice versa; the crystal is irradiated exactly along its $\hat{\vb{z}}$-axis; the two degenerate modes are fully decoupled; and the laser is monochromatic, continuous, and on resonance. Under these assumptions, only one helicity is driven, so we set the occupation of the other to zero. Substituting $Q_{\pm}(t) = \bar{Q}_{\pm} e^{i \omega t}$ into the equation of motion gives the steady-state amplitude $\bar{Q}_{\pm} = \frac{i e Z_{\text{eff}, \pm}^* E_{\pm}}{2 \gamma \omega}$. Since the total energy of a harmonic oscillator in steady state is $\omega^2 \abs{\bar{Q}_{\pm}}^2$, we estimate the occupation number as $n_{\pm} = \frac{\omega}{\hbar} \abs{\bar{Q}_{\pm}}^2 = \frac{\abs*{e Z_{\text{eff}, \pm}^*}^2 \abs{E_{\pm}}^2}{4 \hbar \gamma^2 \omega}.$ A convenient experimental parameter is the laser intensity, related to the field amplitude for circularly polarized light in a medium by $I = c \varepsilon_0 \sqrt{\varepsilon_r} E_{\pm}^2$. Combining these expressions gives our final result:
\begin{equation}
    n^{\pm}_{\nu}(T) = \frac{\abs*{e Z_{\text{eff}, \pm}^*}^2 I}{4 \hbar c \varepsilon_0 \sqrt{\varepsilon}  \gamma_{\nu}(T)^2 \omega_{\nu}}.
    \label{eq:laser_pop}
\end{equation}

Here, we used density functional perturbation theory to compute the Born effective charge tensors $\vb{Z}^*_a$ and the in-plane refractive index, $\sqrt{\varepsilon_{\perp}} \approx 1.53$, on an $8 \times 8 \times 12$ $\Gamma$-centered $k$-mesh. Phonon frequencies and linewidths were computed using $2 \times 2 \times 3$ supercells containing 108 atoms. Because 6493 supercells were required in total, we restricted these calculations to the $\Gamma$-point. We applied the non-analytic term correction (NAC) along the $q$-direction $(0, 0, 1)$, parallel to both the light propagation direction and the crystal $z$-axis and consistently used the ionic mass of thorium-229.

In practice, $^{229}$Th is present only as a dilute dopant, owing to its extreme rarity~\cite{TheThorium229Beeks2021}, substituting for the naturally abundant $^{232}$Th host isotope. We neglect the resulting mass-disorder scattering and its effect on the computed phonon linewidths, a simplification justified on two grounds. First, the isotopic mass mismatch between $^{229}$Th and $^{232}$Th is small. Second, the doping concentration is low, with $ \lesssim \SI{0.1}{\percent}$ being typical of such samples~\cite{Laser_Mossbauer_Hiraki_2025, Continuous_wave_Morawe_2026}. Since mass-disorder scattering rates scale with the mass variance parameter~\cite{Shin1983Isotope, phono3py}, here $g_{\ch{Th}} \lesssim 10^{-7}$, the resulting perturbation to the phonon lifetimes computed in this work is negligible.

\begin{figure}[htpb!]
    \centering
    \includegraphics[width=1.0\linewidth]{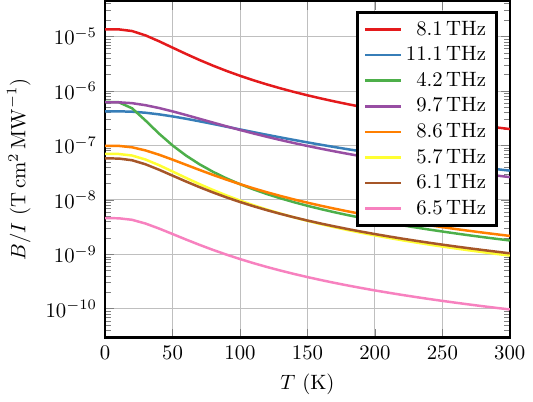}
    \caption{Simulated magnetic flux density induced by a circularly polarized laser as a function of temperature. Since the laser intensity scales linearly, we normalize the $y$-axis by this quantity.}
    \label{fig:laser_B}
\end{figure}

Substituting the expected occupation number, \autoref{eq:laser_pop}, into the magnetic moment expression, \autoref{eq:laser_magmom}, allows us to compute the magnetic field at the thorium nucleus via \autoref{eq:Bfield}, using the DFT-optimized structure. We employed a neighbor distance cutoff of \SI{30}{\angstrom}, excluding the thorium atom at the origin from the summation. \autoref{fig:laser_B} shows the resulting magnetic field response for each degenerate mode.
Among all degenerate modes, the one at \SI{8.1}{\tera\hertz} exhibits the largest magnetic flux at the thorium site across the entire temperature range from \SIrange{0}{300}{\kelvin}. The smallest-frequency mode, at \SI{4.2}{\tera\hertz}, follows with a response more than an order of magnitude weaker at low temperatures. It also shows a favorable temperature dependence, with a pronounced increase in magnetic flux at lower temperatures. This temperature dependence arises from the individual phonon linewidths $\gamma_{\nu}(T)$, as specified in \autoref{eq:laser_pop}. The corresponding shifts in mode frequencies $\omega_{\nu}$ due to thermal expansion are comparatively small and therefore neglected.

\subsection{Applied Temperature Gradient}

\begin{figure}[htbp!]
    \centering
    \includegraphics[width=1.0\linewidth]{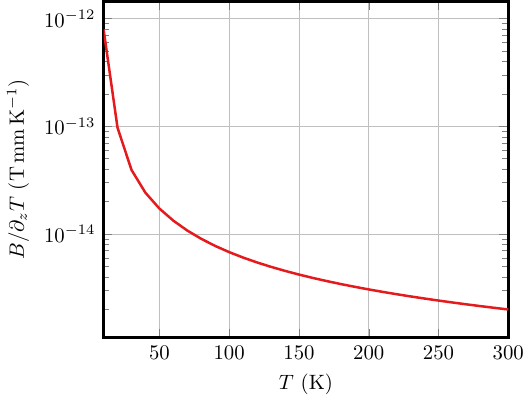}
    \caption{Simulated magnetic flux density induced by an applied temperature gradient as a function of temperature. Since the temperature gradient enters linearly, we normalize the $y$-axis by this quantity.}
    \label{fig:temperature_B}
\end{figure}

The mode magnetic moment per atom, $\vb{M}_{\nu, a}(\vb{q}) = \frac{e \hbar}{2 m_a}  \Im [\vb{e}^*_{\nu, a}(\vb{q}) \times (\vb{Z}^*_a \cdot \vb{e}_{\nu, a}(\vb{q}))]$, is antisymmetric under inversion, $\vb{M}_{\nu}(\vb{q}) = -\vb{M}_{\nu}(-\vb{q})$, while the phonon occupation number is symmetric $n_{\nu}(\vb{q}, T) = n_{\nu}(-\vb{q}, T)$. As a result, the sum of the magnetic moments vanishes at equilibrium. Applying a temperature gradient in non-centrosymmetric crystals breaks this occupational symmetry and gives rise to a net magnetization across the crystal. To compute the resulting deviation from equilibrium occupation, $\delta n = n - n^{(0)}$, where $n^{(0)}(q, T) = \exp(\frac{\hbar \omega_{\nu}(\vb{q})}{k_B T} - 1)^{-1}$, we use the relaxation-time approximation to the linearized Boltzmann transport equation,
\begin{equation}
    \delta n_{\nu}(\vb{q}, T) = -\tau_{\nu}(\vb{q}, T) \vb{v}_{\nu} (\vb{q}) \cdot \grad T(\vb{r}) \pdv{n_{\nu}^{(0)}}{T},
    \label{eq:RTA}
\end{equation}
where $\tau$ is the phonon lifetime, $\vb{v}$ is the group velocity, and $\pdv{n_{\nu}^{(0)}}{T} = \frac{\hbar \omega}{k_B T^2} n_{\nu}^{(0)} (n_{\nu}^{(0)} + 1) $~\cite{Hamada2018}. We apply the temperature gradient along the crystal $z$-axis, $\grad T (\vb{r}) = \partial_z T(\vb{z}) \hat{\vb{z}}$. The net atomic magnetic moment is then obtained by summing over all bands and $\vb{q}$-points, weighted by this occupation-number deviation:
\begin{equation}
    \langle \vb{M}_a \rangle = \frac{e \hbar}{2 m_a} \frac{1}{N_q} \sum_{\vb{q}, \nu} \Im [\vb{e}^*_{\nu, a}(\vb{q}) \times (\vb{Z}^*_a \cdot \vb{e}_{\nu, a}(\vb{q}))] \delta n_{\nu}(\vb{q}, T).
\end{equation}
As in the previous section, we compute the resulting magnetic flux at the thorium site via \autoref{eq:Bfield}, using the same computational parameters and the group velocities obtained from our phonon calculations. We sampled the Brillouin zone with a non-symmetrized $10 \times 10 \times 15$ $q$-mesh (yielding 1500 $q$-points). Results are shown in \autoref{fig:temperature_B}.
The magnetic-field response to a temperature gradient becomes more favorable at lower temperatures.

\begin{figure}[htpb!]
    \centering
    \includegraphics[width=1.0\linewidth]{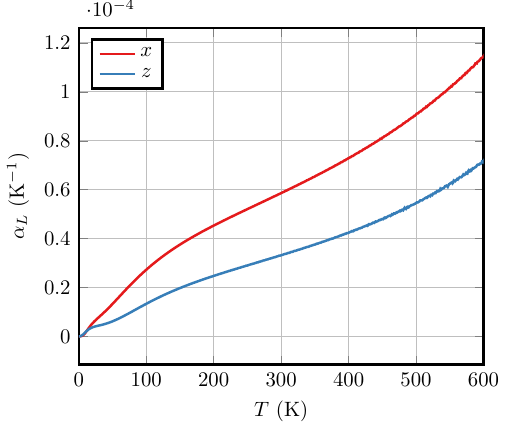}
    \caption{Linear thermal expansion coefficients $\alpha_L$ of \ch{Na_2ThF_6} along its two distinct crystallographic axes, $\alpha_x =\alpha_y$ and $\alpha_z$, obtained by fitting the free energy within the quasi-harmonic approximation.}
    \label{fig:alpha}
\end{figure}

Applying a temperature gradient across the sample also induces inhomogeneous broadening of the probed nuclei, motivating a closer look at the temperature sensitivity of the nuclear transitions. Because the nucleus samples the time-averaged magnetic field along its trajectory, phonon modes are not expected to contribute substantially to line shifts directly; instead, temperature effects arise primarily from thermal expansion of the lattice. The trigonal space group of the crystal permits two independent expansion coefficients, $\alpha_{xx} (= \alpha_{yy})$ and $\alpha_{zz}$. We determine these by minimizing the thermal free energy, $F(T) = U_{\text{el}} + F_{\text{vib}}(T)$, where $U_{\text{el}}$ is the electronic energy from DFT and $F_{\text{vib}}$ is the vibrational free energy obtained from our phonon simulations. At each temperature, we fit $F(x, z) = F_0 + a(x - \varepsilon_{xx})^2 + b(z - \varepsilon_{zz})^2 + cxz$, where $F_0$, $a$, $b$, $c$, $\varepsilon_{xx}$, and $\varepsilon_{zz}$ are fit parameters. We neglect electronic entropy, owing to the material's sizable band gap, as well as the pressure-volume term $pV(T)$. From the resulting thermal strain tensor $\varepsilon(T)$, we compute the linear expansion coefficients as $\alpha_{i}(T) = \pdv{\varepsilon_{ii}}{T}$ (see \autoref{fig:alpha}).

\begin{figure}[htpb!]
    \centering
    \includegraphics[width=1.0\linewidth]{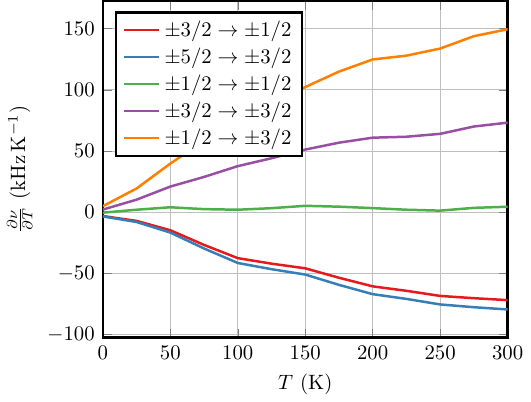}
    \caption{Temperature sensitivity of the allowed nuclear transitions. In the case of the $\pm 1/2 \rightarrow \pm 1/2$ transition, the opposing temperature dependences of the quadrupole and monopole terms nearly cancel, yielding a net sensitivity of $\sim \SI{3}{\kilo \hertz \per \kelvin}$ over the relevant temperature range.}
    \label{fig:sensitivity}
\end{figure}

Finally, we compute the temperature dependence of the nuclear transition energies arising from the isomer shift and quadrupole splitting, $H_{E0} + H_{E2}$ using the temperature-dependent quantities $V_{zz}(T)$ and $\rho_{\text{el}}(0)(T)$. Results are shown in  \autoref{fig:sensitivity}.
We find a fortuitous cancellation between the monopole and quadrupole contributions for the $m_{\text{g}} \rightarrow m_{\text{is}} = \pm 1/2 \rightarrow \pm 1/2$ transition, a feature also identified in Th:CaF$_2$~\cite{PhysRevLett.134.113801}, leading to a temperature sensitivity of only $\pdv{\nu}{T} \sim \SI{3}{\kilo \hertz \per \kelvin}$. This transition is therefore a favorable target for experiments seeking to minimize thermal broadening and maximize signal-to-noise ratio.

\section{Discussion}

\subsection{Theory-Experiment Comparison}

In \autoref{tab:dft_expt}, we compare our simulations (using the thorium-232 mass) to the only experimental reference we could find for \ch{Na_2ThF_6}~\cite{TEIXEIRA2003159}. The disagreement is notable, particularly for the $A_1$ and $A_2$ mode frequencies; these are included for completeness, however, since $A_1$ modes are not IR active and $A_2$ modes do not couple to $x$-$y$ polarized light~\cite{Aroyo:xo5013}, they are not directly relevant to the optical measurements of interest. The $E$ mode frequencies, by contrast, generally match well except for a few high-frequency modes.
\begin{table}[htbp!]
\sisetup{round-mode=places, round-precision=1}
\caption{\label{tab:dft_expt} Comparison between zone-center frequencies in THz obtained numerically in this study with the NAC approaching from $q \parallel z$ and those from an experimental study using Raman and IR-reflectivity measurements~\cite{TEIXEIRA2003159}.}
\begin{ruledtabular}
\begin{tabular}{r r r r r r}
\multicolumn{2}{c}{$A_1$} & \multicolumn{2}{c}{$A_2$ (LO)} & \multicolumn{2}{c}{$E$ (TO)}\\
$\nu_{\text{DFT}}$ & $\nu_{\text{expt.}}$ & $\nu_{\text{DFT}}$ & $\nu_{\text{expt.}}$ & $\nu_{\text{DFT}}$ & $\nu_{\text{expt.}}$ \\
\colrule
\num{2.09101195}    & \num{5.63609821}  & \num{2.28392638}  & \num{2.81804911}  & \num{4.20250333}  & \num{4.52686612}\\
\num{10.28822082}   & \num{8.69398128}  & \num{4.39418975}  & \num{7.88454165}  & \num{5.65067197}  & \num{5.96586991}\\
\num{13.68505681}   & \num{10.43277754} & \num{7.39612003}  & \num{10.61265301} & \num{6.05738023}  & \num{6.50549634}\\
&& \num{8.17107569}     & \num{13.52063986} & \num{6.46017225}  & \num{8.06441712}\\
&& \num{13.20372673}    & \num{14.21016251} & \num{8.05696002}  & \num{8.2442926}\\
&&&& \num{8.61442064}   & \num{9.83319262}\\
&&&& \num{9.69139949}   & \num{11.0623417}\\
&&&& \num{11.05292284}  & \num{13.79045307}
\end{tabular}
\end{ruledtabular}
\sisetup{}
\end{table}

PBEsol phonon frequencies are generally reliable for ionic solids, with a mean absolute error of \SI{4.6}{\percent} in a large high-throughput benchmark, though errors can be substantially larger for materials with strong phonon anharmonicity~\cite{Petretto2018}. Nevertheless, we repeated the phonon calculation of \ch{Na_2 Th F_6} with the r2SCAN functional~\cite{r2scan} and found essentially the same accuracy. Another possible source of discrepancy is DFT's tendency to overestimate dielectric properties~\cite{PhysRevB.89.064305}, where we found $\varepsilon^{\text{DFT}}_{zz} = \num{2.502}$ and $\varepsilon^{\text{DFT}}_{\perp} = \num{2.330}$, compared to the experimental reference of $\varepsilon^{\text{expt.}}_{zz} = \num{2.242}$ and $\varepsilon^{\text{expt.}}_{\perp} = \num{2.272}$, respectively. To test whether thermal expansion could account for this disagreement, we recalculated the phonon spectrum using our computed thermal expansion coefficients for the structure at the experimental conditions of \SI{300}{\kelvin}, but found no substantial improvement.

The experimental data were obtained by combining IR reflectivity and Raman spectroscopy measurements, and surface effects may plausibly affect the former. Clarifying these discrepancies will likely require further experimental work through direct determination of Born effective charges and IR absorption, alongside improved simulations of forces, Born effective charges, and the dielectric properties, as well as inclusion of anharmonic effects, e.g., via velocity autocorrelation functions.

For phonomagnetism, the phonon polarization vectors and linewidths play a more substantial role than frequencies alone, and cannot currently be compared to experiment. Because this study serves primarily as an exploratory investigation aimed at quantitative order-of-magnitude estimates, we leave further improvements to theory-experiment agreement to future work and adopt the DFT values as computed.

\subsection{Comparison of Driving Schemes}

The magnitude of the Zeeman splitting observed in nuclear lines depends on both the magnetic field strength at the nucleus and the intrinsic sensitivity of the relevant nuclear transition to that field. Taking both factors into account, we find that driving degenerate phonon modes with a circularly polarized laser produces substantially larger line splittings than are achievable via an applied temperature gradient. The most sensitive transition can be probed at approximately \SI{1}{\mega\hertz\per\tesla}; conveniently, it also has the highest transition probability, as determined by its Clebsch--Gordan coefficient.

Temperature-gradient-driven schemes suffer from inhomogeneous broadening, further degrading the signal-to-noise ratio. Although we identified one transition that is largely insensitive to temperature broadening, it is also roughly four times less sensitive to magnetic fields. Moreover, laser intensity can be increased by several orders of magnitude to enhance the effective magnetic field, whereas temperature gradients offer comparatively little tunability. Taken together, these factors indicate that detecting a signal from temperature-gradient-driven phonomagnetism would be extremely difficult, if not effectively impossible, using our proposed method.

Since the laser-driven scheme is not tied to the specific thermal-transport properties that motivated our original choice of \ch{Na_2ThF_6}, this conclusion also broadens the pool of viable host crystals as any large-gap stoichiometric crystal hosting degenerate zone-center phonon modes becomes a candidate. Among the 17 large-gap crystals we surveyed, \ch{LiThF_5} and \ch{Li_3ThF_7} stand out in this regard, as they contain Li, whose charge-to-mass ratio substantially exceeds that of either constituent in \ch{Na_2ThF_6} and could therefore yield a stronger phonomagnetic response. A detailed analysis of these materials is left for future work.

\subsection{Experimental Requirements and Laser Systems}

An experimental realization of the circularly polarized laser-driven scheme requires a narrow VUV linewidth to resolve individual nuclear spectral lines spaced only $\sim \si{\mega \hertz}$ apart, as well as high laser intensity, both to generate a measurable magnetic field in the THz regime and to improve the VUV signal-to-noise ratio. Guided by existing thorium laser spectroscopy experiments, we separately analyze continuous-wave (CW) and pulsed laser systems.

\paragraph{Continuous-Wave Lasers} Combining two CW sources, one in the THz and one in the VUV, achieves a true steady state with narrow linewidths in both regimes, but at the cost of low intensity. CW lasers have recently been used for VUV spectroscopy of thorium~\cite{Continuous_wave_Morawe_2026, de2026thorium, huang2026nuclear}. The current state-of-the-art high-intensity CW THz sources are quantum cascade lasers, which emit in the low-THz regime and would need to target the crystal's lowest-frequency degenerate mode~\cite{wang2016, Li2024}.

As an estimate, we consider a laser resonant with the \SI{4.2}{\tera\hertz} mode, delivering \SI{200}{\milli\watt} of power focused onto a \SI{1}{\square\milli\meter} area (\SI{200}{\watt\per\square\centi\meter} intensity). This generates only a weak magnetic field at the thorium nucleus, on the order of \SI{0.1}{\nano\tesla}, corresponding to a line splitting of $\sim \SI{0.1}{\milli\hertz}$, approaching the natural linewidth of the bare thorium transition. Unless CW THz sources with substantially higher power become available, such measurements remain impractical.

\paragraph{Pulsed Lasers}

Pulsed laser systems have also been used for laser spectroscopy of thorium, both with a Ti:sapphire source~\cite{Laser_Mossbauer_Hiraki_2025} and a Yb-fiber source in a frequency comb~\cite{FrequencyRatioZhang2024}. To measure phonomagnetism with pulsed sources, we propose a pump--probe scheme: a THz pump pulse arrives first, populating phonon modes and generating a transient magnetic field at the nucleus, followed by a VUV probe pulse timed to arrive just as the THz pulse ends, reading out the resulting nuclear splitting. Because the two pulses must be precisely synchronized, they should originate from the same source laser, split between the THz and VUV arms~\cite{finneran2015decade}.

A full experimental implementation of this pulsed scheme faces significant technical challenges, including simultaneously achieving high THz conversion efficiency at the required pulse duration and intensity, that lie beyond the scope of this work. In what follows, we restrict ourselves to an order-of-magnitude estimate of the achievable magnetic field.

Because the pulse duration is generally too short for a phonon mode to reach the steady-state occupation of \autoref{eq:laser_pop}, we must account for the transient build-up of the phonon population. For a resonantly driven, damped harmonic oscillator starting from rest, the amplitude approaches its steady-state value as $Q(t) \propto (1 - e^{-\gamma_\nu t})$. For a pulse of duration $\tau$, the resulting occupation number is therefore reduced relative to the CW steady-state value according to
\begin{equation}
    \frac{n_{\text{pulse}}}{n_{\text{CW}}} = \left(1 - e^{-\gamma_\nu \tau}\right)^2.
    \label{eq:pulse_ratio}
\end{equation}
Because this population build-up is exponentially suppressed for short pulses, there exists an optimal pulse length that balances build-up against achievable intensity. To minimize thermal broadening, we consider operation below $\sim \SI{10}{\kelvin}$, where the narrowest linewidths are $\approx \SI{0.02}{\tera\hertz}$; the resulting optimum is $\tau \approx \SI{62}{\pico\second}$ (Fourier-limited bandwidth $\approx \SI{7}{\giga \hertz}$ for Gaussian pulse), corresponding to a population reduction of $n_{\text{pulse}}/n_{\text{CW}} \approx 0.5$ relative to continuous-wave driving.

Since optical-to-THz conversion is typically most efficient at the lower end of the phonon spectrum, we again consider the mode at \SI{4.2}{\tera\hertz}. Reaching a THz conversion efficiency of \SI{0.1}{\percent} at this frequency with a \SI{62}{\pico\second}, narrowband pulse is a nontrivial engineering task; however, comparable efficiencies have already been demonstrated in related regimes~\cite{Jolly2019, buchmann2020high, Seo2022, wangkangzheng, arxiv.2503.23411Nasi, Brekhov:25}, and we adopt this value for our order-of-magnitude estimate.

Considering the thorium laser spectroscopy experiments discussed above, a beam focused onto \SI{1}{\square \milli \meter} yields a pulse energy of $\sim \SI{1}{\joule \per \square \centi \meter}$~\cite{FrequencyRatioZhang2024, Laser_Mossbauer_Hiraki_2025}. Combined with the estimated pulse duration and THz-generation efficiency, this corresponds to an expected intensity of \SI{16}{\mega\watt\per\square\centi\meter}. The resulting order-of-magnitude estimate yields an effective magnetic field of $\sim\SI{5}{\micro\tesla}$ and a corresponding splitting of $\sim\SI{5}{\hertz}$.

\subsection{Sensitivity and Outlook}

The linewidth and practical stability of the VUV probe laser ultimately determine the magnetic-field sensitivity achievable with this scheme. The shot-noise-limited frequency resolution is given by
\begin{equation}
\delta\nu_{\min} \approx \Gamma/\sqrt{N},
\label{eq:shotnoise}
\end{equation}
where $\Gamma$ is the laser linewidth and $N$ is the number of photons collected per data point. Taking $N \approx 10^6$, consistent with previous measurements of this transition using a frequency comb~\cite{FrequencyRatioZhang2024}, we obtain $\delta\nu_{\min} \approx \Gamma/1000$.

Although a stoichiometric crystal has not yet been produced, we estimate this intrinsic linewidth to lie between $\sim \SI{1.5}{\kilo\hertz}$ and $\sim \SI{25}{\kilo\hertz}$. The lower bound derives from an upper estimate of the $^{19}$F spin-bath broadening~\cite{PerformanceOfKazako2012}, while the upper bound comes from extrapolating defect-induced strain linewidths to zero concentration~\cite{Frequency_repro_Ooi_T_2026}; an intermediate estimate of $\sim \SI{5}{\kilo\hertz}$ arises from a similar extrapolation to zero EFG-induced broadening~\cite{Li2026}. Both extrapolations rely on measurements from three doping concentrations in \ch{Th:CaF_2}, and although the spin-bath broadening was estimated specifically for \ch{Th:CaF_2}, the similar Th-F distance in other fluorides suggests this lower bound may hold more generally.

Given \autoref{eq:shotnoise}, the shot-noise-limited resolution thus ranges from $\sim \SIrange{1.5}{25}{\hertz}$ after long integration times, such as those employed in clock measurements~\cite{de2026thorium}. Such precision is enabled by frequency-comb techniques, which have already achieved highly accurate thorium spectroscopy via single-tooth excitation~\cite{FrequencyRatioZhang2024, Frequency_repro_Ooi_T_2026}. Notably, the same comb's pulse train can also directly generate the high-intensity THz pulses required to drive the phonomagnetic response~\cite{finneran2015decade}. The pulsed THz laser is expected to produce a magnetic field inducing a $\sim \SI{5}{\hertz}$ shift; whether this signal is resolvable therefore depends on where the linewidth falls within our estimated range. If the linewidth lies near the lower end of this range, comparable to or below the projected shift, detection should be feasible, albeit challenging. If instead the linewidth approaches the upper bound, the signal would remain unresolved with the present scheme. A definitive assessment, however, requires first characterizing the linewidth in a stoichiometric crystal.

Because measured phonon magnetic moments have consistently exceeded theoretical predictions by 3--4 orders of magnitude~\cite{Chaudhary24Giant, Axial_phono_mag_Shabal_2025, Chen25Geometric, Chen25Gauge}, the values used here should be regarded as a conservative lower bound on the effect's strength; we nonetheless design our proposed experiment around these lower-bound estimates. As a result, since the predicted phonomagnetic field in \ch{Na_2ThF_6} is weak, our proposed measurement sits at the borderline of the sensitivity needed to test the theory. However, this is not a shortcoming of the proposal: each possible outcome of the experiment (null, confirming, or exceeding the theoretical estimate) would constitute a meaningful and interpretable result. Taken together, these considerations support the conclusion that a circularly polarized pump--probe scheme, while experimentally demanding, offers a promising and direct route toward measuring phonomagnetic fields in \ch{Na_2 Th F_6}.

\section{Conclusion}

We propose using the low-energy nuclear transition of \ch{^{229}Th} in a stoichiometric crystal as a microscopic, in-situ probe of the magnetic fields generated by circularly polarized phonons. This approach offers two key advantages: it allows the field to be probed directly inside the host crystal without requiring doping, and it benefits from a low intrinsic magnetic background. We considered two mechanisms for generating a phonomagnetic field: driving degenerate optical modes at the $\Gamma$-point with a THz laser, and applying a temperature gradient across the material.

\ch{Na_2ThF_6} satisfies all the criteria required for this scheme: it hosts degenerate $E$-modes at the zone center, is non-centrosymmetric, and is transparent at the nuclear clock transition frequency. Our DFT study of this material indicates that the temperature-gradient-driven phonon magnetism is too weak to yield an observable response using current methods. In contrast, a laser-driven scheme can cause nuclear-phonomagnetic line splittings that, while small, approach the projected shot-noise floor when combined with high-intensity, narrowband pulses of a pump--probe system.

Because the predicted signal is weak and lies close to this sensitivity limit, the experiment is well suited to distinguishing between physically distinct outcomes: No observable splitting would be consistent with a phonomagnetic field in \ch{Na_2ThF_6} smaller than predicted; a splitting matching the predicted value would confirm the theory under clean experimental conditions; and a substantially larger splitting would identify \ch{Na_2ThF_6} as another material with an underestimated phonon magnetic moment. Realizing this potential in practice, however, will require overcoming substantial experimental challenges, including isolating the phonomagnetic signal from competing systematic effects, and demonstrating sufficient sensitivity at the Hz level.  Future work may extend this analysis to other large-gap fluorides, such as \ch{LiThF_5} and \ch{Li_3ThF_7}, whose higher Li charge-to-mass ratio could yield a stronger phonomagnetic response.


\begin{acknowledgments}
We thank M. Geilhufe and D. Juraschek for stimulating conversations that motivated aspects of this study. This work has been funded by the European Research Council (ERC) under the European Union’s Horizon 2020 research and innovation programme (Grant Agreement No. 856415 and 101087184) and the Austrian Science Fund (FWF) [Grant DOI: 10.55776/J4834, 10.55776/PIN9526523]. The computational results have been achieved using the Austrian Scientific Computing (ASC) infrastructure. During preparation of this work, we used Claude Sonnet 5, Opus 5.5, Fable 5, and Fable 5.1 to aid comprehension of derivations in previously published theory and to aid in interpreting select conclusions and assumptions. AI-assisted content was incorporated only where corroborated by our own analysis. We take full responsibility for the content of the manuscript.
\end{acknowledgments}


\bibliography{refs}
\end{document}